\documentclass[conference]{IEEEtran}
\IEEEoverridecommandlockouts
\usepackage{amsthm}
\usepackage{amsmath}
\usepackage{latexsym}
\usepackage{graphicx}
\usepackage{bbding}
\usepackage{indentfirst}
\usepackage{algorithm,algorithmic}
\usepackage{setspace}
\usepackage{float}
\usepackage{epstopdf}
\usepackage{amssymb}
\usepackage{amsfonts}
\usepackage{enumerate}
\usepackage{multicol}
\usepackage{color}
\usepackage{diagbox}
\usepackage{bm}
\usepackage{amssymb}
\usepackage{stfloats}
\usepackage{epstopdf}
\usepackage{threeparttable}
\usepackage{epstopdf}
\usepackage{threeparttable}
\usepackage{subfigure}
\usepackage{makecell}
\usepackage{cite}

\begin{document}
	
	\title{Hybrid-Field 6D Movable Antenna for Terahertz Communications: Channel Modeling and Estimation}
	
		\author{\IEEEauthorblockN{Xiaodan Shao$\IEEEauthorrefmark{1}$, Yixiao Zhang$\IEEEauthorrefmark{1}$, Shisheng Hu$\IEEEauthorrefmark{1}$, Zhixuan Tang$\IEEEauthorrefmark{1}$,  Mingcheng He$\IEEEauthorrefmark{1}$, Xinyu Huang$\IEEEauthorrefmark{1}$,
			Weihua Zhuang$\IEEEauthorrefmark{1}$, \\and Xuemin (Sherman) Shen$\IEEEauthorrefmark{1}$}
		\IEEEauthorblockA{$\IEEEauthorrefmark{1}$Department of Electrical and Computer Engineering, University of Waterloo, Canada\\
			E-mails: \{x6shao, y3549zha, s97hu, z32tang, m64he, x357huan, wzhuang, sshen\}@uwaterloo.ca}\vspace{-23pt}	
	}\maketitle

	\IEEEpeerreviewmaketitle
	\begin{abstract}
In this work, we study a six-dimensional movable antenna (6DMA)-enhanced Terahertz (THz) network that supports a large number of users with a few antennas by controlling the three-dimensional (3D) positions and 3D rotations of antenna surfaces/subarrays at the base station (BS).
		However, the short wavelength of THz signals combined with a large 6DMA movement range extends the near-field region. As a result, a user can be in the far-field region relative to the antennas on one 6DMA surface, while simultaneously residing in the near-field region relative to other 6DMA surfaces. Moreover, 6DMA THz channel estimation suffers from increased computational complexity and pilot overhead due to uneven power distribution across the large number of candidate position-rotation pairs, as well as the limited number of radio frequency (RF) chains in THz bands. To address these issues, we propose an
		efficient hybrid-field generalized 6DMA THz channel model, which accounts for planar wave propagation within individual 6DMA surfaces and spherical waves among different 6DMA surfaces. Furthermore, we propose a low-overhead channel estimation algorithm that leverages directional sparsity to construct a complete channel map for all potential antenna position-rotation pairs.
		Numerical results show that the proposed hybrid-field channel model achieves a sum rate close to that of the ground-truth near-field channel model and confirm that the channel estimation method yields accurate results with low complexity.
	\end{abstract}

\section{Introduction}
Terahertz (THz) wireless communications, operating in the 0.1 to 10 THz band, are regarded as a promising solution to support ultra-broadband for future 6G networks \cite{thzw}. Nevertheless, THz signals face severe free-space loss and significant non-line-of-sight (NLoS) attenuation. These factors limit transmission reliability in obstructed environments. To address this issue, ultra-massive multiple-input multiple-output (MIMO) systems are employed in the THz band by deploying large-scale antenna arrays. Using hundreds or thousands of antennas helps provide sufficient array gain to mitigate propagation losses \cite{10540249,9540383,9448777,9432908,9266124,9146533,8612905,8937497,9973387,10115008,9844229,11406144,zhai2026does}
. However, increasing the number of antennas essentially comes at the expense of increased hardware cost. Moreover, such fixed-position MIMO systems lack the flexibility to adjust antenna configurations based on the actual spatial distribution of user channels.

Recently, six-dimensional movable antenna (6DMA) has been proposed to increase MIMO system capacity without requiring additional antennas, by fully exploiting spatial variations of the wireless channel at transceivers \cite{shao20246d, 6dma_dis}. This technique leverages the adaptability of antenna positions and orientations in three-dimensional (3D) space at transceivers to adaptively allocate antenna resources based on the spatial distribution of channels. Building on these advancements, an unmanned aerial vehicle (UAV)-enabled 6DMA architecture was explored in \cite{chenzhi6}, where a 6DMA surface is installed on a UAV and joint antenna deployment and beamforming optimization are investigated. Furthermore, emerging research paradigms of 6DMA-aided intelligent reflecting surfaces (IRSs) have been studied in \cite{qingmma}. In addition, coupler position and rotation optimization for flexible coupler antennas (FCAs) was studied in~\cite{FC,FC1,shao2026coupler,FCjstsp,FCATWC,RCATWC}, where controllable mutual coupling was exploited to enhance antenna reconfigurability with a reduced number of radio frequency (RF) chains.   

\begin{figure}[t!]
			\setlength{\abovecaptionskip}{-3pt}
		\setlength{\belowcaptionskip}{-15pt}
		\centering
		\setlength{\abovecaptionskip}{0.cm}
		\includegraphics[width=3.60in]{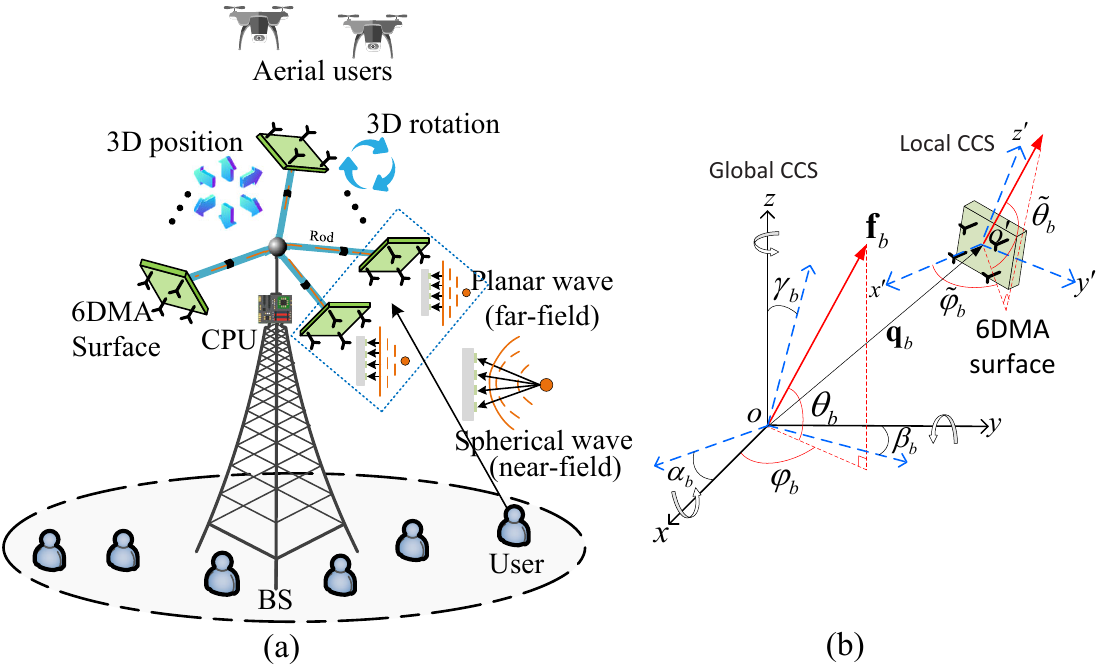}
		\caption{(a) A typical hybrid-field propagation environment in a 6DMA THz system; (b) The geometry of hybrid-field channel model.}
		\label{practical_scenario}
			\vspace{-0.58cm}
	\end{figure}

Motivated by the advancements of 6DMA \cite{wang2026flexible,li2025ai,li2025flexible,11148174,shi2025throughput}, we study a THz wireless communication 
system enhanced with multiple 6DMAs at the base station (BS) (see Fig. \ref{practical_scenario}(a)). The 6DMA THz system faces two main challenges. First, although the existing work on 6DMA channel modeling considers only the far-field propagation \cite{shao20246d, 6dma_dis,jiang2025statistical,jiang2026multi,9724202,10740590,free6DMA,shao2025tutorial,6DMA_JSTSP,passive6DMA,6dmasensing,10945745,near,IPA,10969546,11493774,11314850,11371617,11181061,11007737,11454630,11432796}, the near-field region expands with increased dimensions of antenna movement regions and higher carrier frequencies. However, applying the near-field channel model to 6DMA leads to an exponential increase in the number of parameters, which is proportional to the massive number of candidate positions and possible rotations. 
In practice, 6DMA surfaces may be spaced significantly apart from each other, which results in users being positioned in the near-field region relative to different 6DMA surfaces. On the other hand, antennas on each 6DMA surface are positioned relatively close to one another, placing users in the far-field region of these antennas (see Fig. \ref{practical_scenario}(a)). To effectively capture this hybrid-field channel feature, we propose a hybrid-field 6DMA THz channel model, which includes the far-field and near-field channel models as special cases. This approach not only achieves better accuracy than the traditional far-field channel model but also requires fewer channel parameters than the near-field channel model. 

Second, to achieve optimal 6DMA performance, it is imperative to acquire accurate channel state information (CSI) linking every possible candidate antenna configuration with its user. However, 6DMA must acquire CSI in a continuous space with a vast number of candidate antenna positions and/or rotations, which results in high computational complexity and pilot overhead. Furthermore, since hybrid MIMO structures in THz systems employ far fewer radio frequency (RF) chains than antenna elements CSI estimation in THz systems must recover a high-dimensional channel corresponding to the number of candidate antenna position-rotation pairs from the limited signals available at the RF chains. To overcome this challenge, we propose an efficient channel estimation algorithm to enable the inference of channel information for any unmeasured position-rotation pair by considering uneven channel power distribution properties through the concept of directional sparsity of the 6DMA THz channel. 

\emph{Notations}: Notation \( \mathbb{E}[\cdot] \) denotes the expected value of a random variable, \( \left \| \cdot \right \|_2 \) denotes the Euclidean norm, \(|\cdot|\) denotes the absolute value, \( |\cdot|_{\mathrm{c}} \) denotes the cardinality of a set, and $\text{blkdiag}(\cdot)$ denotes the block-diagonal operator that arranges its input matrices along the main diagonal of a larger matrix.
	\section{System Model}
	\subsection{6DMA System Model}
	As shown in Fig. \ref{practical_scenario}(a), we consider a narrow-band THz system, in which the 6DMA-enabled BS has multiple 6DMA surfaces, denoted by  \(\mathcal{B} = \{1, 2, \ldots, B\}\), and serves \(K\) single-antenna users \cite{shao20246d, 6dma_dis}. Each 6DMA surface is set up as a uniform planar array (UPA) with \(N \geq 1\) antennas, which are represented by \(\mathcal{N} = \{1, 2, \ldots, N\}\). Each 6DMA surface is connected to one RF-chain, and a surface-based hybrid beamforming structure is configured at the BS side. These surfaces are lined to a central processing unit (CPU) via extendable and rotatable rods, which enable flexible adjustments in their 3D positions and 3D rotations. 
	
	The position and rotation of the \(b\)-th 6DMA surface, where \(b \in \mathcal{B}\), are determined by six parameters: \(\mathbf{q}_b\) for 3D position and \(\mathbf{u}_b\) for 3D rotation (see Fig. \ref{practical_scenario}(b)). The values for these parameters are represented by \cite{shao20246d, 6dma_dis}
	\begin{align}\label{bb}
		\mathbf{q}_b=[x_b,y_b,z_b]^T\in\mathcal{C},~	\mathbf{u}_b=[\alpha_b,\beta_b,\gamma_b]^T,
	\end{align}
	where $\mathcal{C}\in\mathbb{R}^3$ denotes the 3D space within which the 6DMA surfaces can be positioned/rotated. Here, \( x_b \), \( y_b \), and \( z_b \) denote the 3D coordinates of the center of the \( b \)-th 6DMA in a global Cartesian coordinate system (CCS) \( o\text{-}xyz \), where the origin, \( o \), is set at the BS's reference position. The angles, \( \alpha_b \), \( \beta_b \), and \( \gamma_b \), each range from 0 to \( 2\pi \) and represent rotation angles around the $x$-axis, $y$-axis, and $z$-axis, respectively.
	The rotation matrix, \( \mathbf{R}(\mathbf{u}_b) \), with respect to \( \mathbf{u}_b \) is given by
\begin{align}\label{R}
	&\!\mathbf{R}(\mathbf{u}_b)\!=\!\begin{bmatrix}
		c_{\beta_b}c_{\gamma_b} & c_{\beta_b}s_{\gamma_b} & -s_{\beta_b} \\
		s_{\beta_b}s_{\alpha_b}c_{\gamma_b}-c_{\alpha_b}s_{\gamma_b} & s_{\beta_b}s_{\alpha_b}s_{\gamma_b}+c_{\alpha_b}c_{\gamma_b} & c_{\beta_b}s_{\alpha_b} \\
		c_{\alpha_b}s_{\beta_b}c_{\gamma_b}+s_{\alpha_b}s_{\gamma_b} & c_{\alpha_b}s_{\beta_b}s_{\gamma_b}-s_{\alpha_b}c_{\gamma_b} &c_{\alpha_b}c_{\beta_b} \\
	\end{bmatrix},\!\!
\end{align}
	where $c_x = \cos(x)$ and $s_x = \sin(x)$.	
	Each 6DMA surface has its local coordinate system denoted as \( o' \text{-} x'y'z'\), with \(o'\) as the origin situated at the center of the surface. Let \( \bar{\mathbf{r}}_{n}\) represent the position of the \( n \)-th antenna on the \(b\)-th 6DMA surface in its local CCS. Then, the global position of the \( n \)-th antenna on the \( b \)-th 6DMA surface in the global CCS is expressed as \cite{shao20246d, 6dma_dis}
	\begin{align}\label{nwq}
		\mathbf{r}_{b,n}(\mathbf{q}_b,\mathbf{u}_b)=\mathbf{q}_b+\mathbf{R}
		(\mathbf{u}_b)\bar{\mathbf{r}}_{n},~n\in\mathcal{N},~b \in\mathcal{B}.
	\end{align}
	\subsection{6DMA THz Channel Model} 
Due to the severe attenuation induced by scattering in the THz band, we neglect the NLoS components and only consider the dominant line-of-sight (LoS) channel between the location of any user and the 6DMA-enabled BS. We begin by presenting the far-field and near-field 6DMA channel models, and subsequently introduce the proposed hybrid-field channel model.
\subsubsection{Far-Field 6DMA Channel Model}
Let $\phi\in[-\pi,\pi]$ and $\theta\in[-\pi/2,\pi/2]$ represent the azimuth and elevation angles, respectively, of a signal from any user with regard to BS center $o$. The unit-length direction-of-arrival (DOA) vector corresponding to direction $(\theta, \phi)$ is given by 
\begin{align}\label{KM}
	\hat{\mathbf{f}}=[\cos(\theta)\cos(\phi), \cos(\theta)\sin(\phi), \sin(\theta)]^T.
\end{align}
By combining \eqref{nwq} and \eqref{KM}, the steering vector for the \(b\)-th 6DMA surface can be expressed as \cite{shao20246d,6dma_dis}
\begin{align}\label{gen}
	\hat{\mathbf{a}}(\mathbf{q}_b,\mathbf{u}_b)=& \left[e^{-j\frac{2\pi}{\lambda}
		\hat{\mathbf{f}}^T\mathbf{r}_{b,1}(\mathbf{q}_b,\mathbf{u}_b)},
	\cdots, e^{-j\frac{2\pi}{\lambda}\hat{\mathbf{f}}^T
		\mathbf{r}_{b,N}(\mathbf{q}_b,\mathbf{u}_b)}\right]^T,
\end{align}
where $\lambda=\frac{c}{f_c}$ denotes the carrier wavelength with $f_c$ and $c$ representing the center carrier frequency and the speed of light, respectively. Then, the far-field channel between any user's location and the 6DMA-enabled BS is expressed as
\begin{align}\label{uk}
	&\!\!\mathbf{h}^{\mathrm{far}}(\mathbf{q},\mathbf{u})\!
	=\nonumber\\
	&\!\!\nu
	e^{-j\frac{2\pi d}{\lambda}}\!\left[\!\sqrt{\hat{g}(\mathbf{u}_1)}
	\hat{\mathbf{a}}^T(\mathbf{q}_1,\mathbf{u}_1),
	\!\cdots\!,\sqrt{\hat{g}(\mathbf{u}_B)}\hat{\mathbf{a}}^T(\mathbf{q}_B,\mathbf{u}_B)
	\!\right]^T,\!\!
\end{align}
with $
{\mathbf{q}}=[\mathbf{q}_1^T,\cdots,\mathbf{q}_B
^T]^T\in \mathbb{R}^{3B\times 1}$ and $
\mathbf{u}=[\mathbf{u}_1^T,\cdots,\mathbf{u}_B^T]^T\in \mathbb{R}^{3B\times 1}$.
In the above, $d$ denotes the distance between any user's location and the reference position of the BS, $\nu$ denotes the path gain, and $\hat{g}(\mathbf{u}_b)$ denote the effective antenna gain for the $b$-th 6DMA surface in linear scale. To define $\hat{g}(\mathbf{u}_b)$, we project the DOA vector, \( \hat{\mathbf{f}} \), onto the local CCS of the \( b \)-th 6DMA surface as
\begin{align}
	\hat{\mathbf{f}}_b=\mathbf{R}^{-1}(\mathbf{u}_b)
	\hat{\mathbf{f}}.
\end{align}
Then, we represent $\hat{\mathbf{f}}$ in the spherical coordinate system as
\begin{align}\label{KM3}
	\hat{\mathbf{f}}_b=[\cos(\hat{\theta}_{b})\cos(\hat{\phi}_{b}), \cos(\hat{\theta}_{b})\sin(\hat{\phi}_{b}), \sin(\hat{\theta}_{b})]^T,
\end{align}
where $\hat{\theta}_{b}$ and $\hat{\phi}_{b}$ represent the corresponding DOAs in the local CCS. Next, $\hat{g}(\mathbf{u}_b)$ can be represented as 
\begin{align}\label{gm}
	\hat{g}(\mathbf{u}_b)=10^{\frac{A(\hat{\theta}_{b}, \hat{\phi}_{b})}{10}},
\end{align}
where \( A(\hat{\theta}_{b}, \hat{\phi}_{b}) \) denotes the effective gain of each antenna on the \( b \)-th 6DMA surface, measured in decibels isotropic (dBi). 


\subsubsection{Near-Field 6DMA Channel Model}
In the near-field 6DMA channel model, we define $\phi_{b,n}\in[-\pi,\pi]$ and $\theta_{b,n}\in[-\pi/2,\pi/2]$ as the azimuth and elevation angles, respectively, of a signal from any user with respect to the $n$-th antenna on the $b$-th 6DMA surface. Then, the near-field channel model is given by
\begin{align}
	& \mathbf{h}^{\mathrm{near}}(\mathbf{q},\mathbf{u})	= [\nu_{1,1}{\sqrt{\bar{g}(\mathbf{q}_1,\mathbf{u}_1)}}
	^{-j\frac{2\pi}{\lambda}
		\bar{d}_{1,1}},
	\cdots, \nu_{b,n}{\sqrt{\bar{g}(\mathbf{q}_b,\mathbf{u}_b)}}\nonumber\\
	&
	e^{-j\frac{2\pi}{\lambda}
		\bar{d}_{b,n}},\cdots,	\nu_{B,N}{\sqrt{\bar{g}(\mathbf{q}_B,\mathbf{u}_B)}}
	e^{-j\frac{2\pi}{\lambda}
		\bar{d}_{B,N}}]^T, \label{hnn}
\end{align}
where \( \bar{d}_{b,n} \) represents the distance between any user and the \(n\)-th antenna on the \(b\)-th 6DMA surface at the BS, $\nu_{b,n}$ denotes the corresponding path gain, and the antenna gain is given by \( \bar{g}(\mathbf{q}_b, \mathbf{u}_b) = 10^{\frac{A(\bar{\theta}_{b,n}, \bar{\phi}_{b,n})}{10}} \).  Similar to \eqref{gm}, the angles \( \bar{\theta}_{b,n} \) and \( \bar{\phi}_{b,n} \) can be obtained by projecting DOA vector \( \bar{\mathbf{f}}_{b,n}\!=\![\cos(\theta_{b,n})\cos(\phi_{b,n}), \cos(\theta_{b,n})\sin(\phi_{b,n}),\sin(\theta_{b,n})]^T \) onto the local CCS of the \( b \)-th 6DMA surface through $\mathbf{R}(\mathbf{u}_b)^T\bar{\mathbf{f}}_{b,n}$.

\subsubsection{Hybrid-Field 6DMA Channel Model}
The proposed hybrid-field 6DMA model 
accounts for far-field channel model within one 6DMA surface and near-field channel model among 6DMA surfaces by assigning a unique signal transmission angle to every surface. 
Specifically, we let $\phi_{b}\in[-\pi,\pi]$ and $\theta_{b}\in[-\pi/2,\pi/2]$ denote the azimuth and elevation angles, respectively, of a signal from any user with respect to (w.r.t.) the center of the $b$-th 6DMA surface (see Fig. \ref{practical_scenario}(b)).
The unit-length DOA vector corresponding to direction $(\theta_b, \phi_b)$ is given by
\begin{align}
	{\mathbf{f}}_{b}=[\cos(\theta_{b})\cos(\phi_{b}), \cos(\theta_{b})\sin(\phi_{b}), \sin(\theta_{b})]^T.\label{KMh}
\end{align}
Consequently, the steering vector of the \( b \)-th 6DMA surface is given by
\begin{align}
	\!\!\mathbf{a}(\mathbf{q}_b,\!\mathbf{u}_b)\!= \!\left[\!e^{-j\frac{2\pi}{\lambda}
		{\mathbf{f}}_{b}^T\mathbf{r}_{b,1}(\mathbf{q}_b,\mathbf{u}_b)},
	\!\cdots\!, e^{-j\frac{2\pi}{\lambda}{\mathbf{f}}_{b}^T
		\mathbf{r}_{b,N}(\mathbf{q}_b,\mathbf{u}_b)}
	\!\right]^T, \!\!\!\label{genhtt}
\end{align}

Based on \eqref{KMh} and \eqref{genhtt}, we construct the following hybrid-field channel model between any user's location and the BS\footnote{The hybrid-field channel model can be extended to a general multipath channel, where each signal path can be similarly modeled as the LoS path in \eqref{ap}.}:
\begin{align}
	&\!\!\!\!\!\!\!\mathbf{h}(\mathbf{q},\mathbf{u})=
	\nu[{ \sqrt{g(\mathbf{q}_1,\mathbf{u}_1)}e^{-j\frac{2\pi}{\lambda}
			d_{1}}}
	\mathbf{a}^T(\mathbf{q}_1,
	\mathbf{u}_1),\cdots,\nonumber\\
	&{ \sqrt{g(\mathbf{q}_B,\mathbf{u}_B)}e^{-j\frac{2\pi}{\lambda}
			d_{B}}}
	\mathbf{a}^T(\mathbf{q}_B,\mathbf{u}_B)
	]^T\in\mathbb{C}^{NB\times 1},\label{ap}
\end{align}
where $d_{b}$ denotes the distance between any user's location and the $b$-th 6DMA surface, and the corresponding antenna gain is given by \( g(\mathbf{q}_b,\mathbf{u}_b)=10^{\frac{A(\tilde{\theta}_{b}, \tilde{\phi}_{b})}{10}} \).  Similar to \eqref{gm}, the angles \( \tilde{\theta}_{b} \) and \( \tilde{\phi}_{b} \) can be obtained by projecting the DOA vector \( {\mathbf{f}}_{b}\) in \eqref{KMh} onto the local CCS of the \( b \)-th 6DMA surface adopting $
\mathbf{R}
(\mathbf{u}_b)^T{\mathbf{f}}_{b}$ (see Fig. \ref{practical_scenario}(b)).
	
\textbf{Remark 1}:
When $B=1$, the proposed hybrid-field 6DMA channel model reduces to the conventional far-field 6DMA model in \cite{shao20246d}; when $N=1$, the hybrid-field model effectively reduces to the near-field 6DMA channel model, which has not been studied in the literature. 

\subsection{Signal Model}
Before downlink data transmission, the users first transmit orthogonal training pilots to the BS over $T$ time slots for CSI acquisition. Therefore, channel estimation
can be performed independently from each user and we can consider an arbitrary user without loss of generality.
During this phase, all 6DMA surfaces traverse a small number of
$M$ different position-rotation pairs, denoted by $\{\mathbf{q}_m,\mathbf{u}_m\}_{m=1}^M$, to collect channel
measurements for reconstructing the channel information for all available antenna locations and rotations.

Since each 6DMA surface has its own RF chain and analog beamforming module, the training procedure can be completed surface-wise. Specifically, at the $m$-th 6DMA position-rotation pair, the received signal is given by
\begin{align}
	y_{m,t}=\mathbf{w}_{m,t}^H\mathbf{h}_m(\mathbf{q}_m,\mathbf{u}_m)s_{t}+
	\mathbf{w}_{m,t}^H\mathbf{z}_{m,t},
\end{align}
where $\mathbf{h}_m(\mathbf{q}_m,\mathbf{u}_m)=\nu{ \sqrt{g(\mathbf{q}_m,\mathbf{u}_m)}e^{-j\frac{2\pi}{\lambda}
		d_{m}}}
\mathbf{a}^T(\mathbf{q}_m,
\mathbf{u}_m)$ denotes the channel between a user and the $m$-th 6DMA position-rotation pair, $s_{t}$ is the transmitted pilot in time slot $t$, $\mathbf{w}_{m,t}\in \mathbb{C}^{N \times 1}$ denotes the analog combining vector at the $t$-th time slot satisfying the constant modulus constraint, and 
 \(\mathbf{z}_{m,t}\) denotes the received Gaussian noise vector with zero mean and covariance matrix $\sigma^2 \mathbf{I}_{N}$.
Collecting 	$y_{m,t}$ over $T$ time slots, we obtain
\begin{align}
\!\!\mathbf{y}_m \!=\! \mathbf{W}_m \mathbf{h}_m(\mathbf{q}_m,\mathbf{u}_m) \!+\! \mathbf{z}_m \in \mathbb{C}^{T \times 1}, \label{mo}
\end{align}
for $m\!\in\!\mathcal{M}\!=\!\{1,2,\cdots, M\}$, where we assume $s_{t}\!=\!1$ for $t=1,2,\cdots,T$ without loss
of generality. Here, 
 $\mathbf{W}_m=[\mathbf{w}_{m,1},\mathbf{w}_{m,2},\cdots,\mathbf{w}_{m,T}]^T \in \mathbb{C}^{T \times N}$ represents the analog combining matrix, and $\mathbf{z}_m=[\mathbf{w}_{m,1}^T\mathbf{z}_{m,1},\cdots,\mathbf{w}_{m,T}^T\mathbf{z}_{m,T}] $.
\section{Channel Estimation}
In this section, we present the proposed channel estimation method to efficiently estimate the 6DMA THz channel with affordable complexity. Note that for, \(b \in \mathcal{B}\), parameters \(d_b\) and \((\phi_b, \theta_b)\) in \eqref{ap} can be derived from \(d\) and \((\phi, \theta)\), respectively, according to the channel's geometric relationships. Therefore, the hybrid-field 6DMA  channel becomes a function of path gain \(\nu\), user-BS distance \(d\), and the signal's azimuth and elevation angles, \(\phi\) and \(\theta\), relative to the BS center. Therefore, by accurately determining \(\nu\), \(d\), \(\phi\), and \(\theta\) from the received signal, the complete 6DMA channel between users and all possible 6DMA positions and rotations at the BS can be reconstructed. 

Building on this foundation, the proposed scheme leverages the directional sparsity property (discussed later in Section III-A) of 6DMA channels and is implemented in two steps. First, all 6DMA surfaces move over a total of $M$ different position-rotation pairs to collect channel measurements, and each 6DMA position‑rotation pair (i.e., its surface) produces an independent coarse estimate. This step offers scalability and low computational complexity by treating each surface independently. Then, a refined parameter estimation is carried out by adopting those 6DMA position-rotation pairs that exhibit strong received signals. This collaborative approach improves channel reconstruction accuracy by exploiting the uneven channel power distribution resulting from directional sparsity. In the following, we introduce the details of the designed channel estimation algorithm.
\subsection{Unequal Channel Power from Directional Sparsity}
The 6DMA THz system exhibits an uneven channel power distribution (i.e., spatial non-stationarities) across its surfaces due to an inherent directional sparsity property (see Fig. \ref{heatmap} in Section IV). Specifically, due to the antenna directivity of 6DMA, each user may have channels with significant gains only for a small subset of all possible 6DMA position-rotation pairs indexed by $\mathcal{W}\subseteq \mathcal{M}$
(i.e., antenna gain $g(\mathbf{q}_m,\mathbf{u}_m)\neq 0$ for all $ m\in \mathcal{W}$). In contrast, for the remaining candidate position-rotation pairs, which may either face in the opposite direction of the user or be blocked by obstacles towards the user, the effective channel gains of the user are significantly weaker and thus can be ignored (i.e.,  $g(\mathbf{q}_m,\mathbf{u}_m)=0$ for all $m\in \mathcal{W}^{\mathrm{c}}$, where $\mathcal{W} \cup \mathcal{W}^{\mathrm{c}} = \mathcal{M}$ and $ \mathcal{W} \cap \mathcal{W}^{\mathrm{c}} = \emptyset
$). We refer to this characteristic as {\textit{directional sparsity}}, which is unique to 6DMA system \cite{shao20246d}. For example, in Fig. \ref{practical_scenario}(a), a user located on the ground can establish a strong channel with the 6DMA surface directed toward it, whereas the channel with the 6DMA surface directed toward the sky is much weaker and can be approximated as zero.

Such directional sparsity causes negligible signal power to be observed at some 6DMA position-rotation pairs for a given user. This uneven power distribution, along with the high dimensionality of candidate antenna position-rotation pairs, has significantly harmful impacts on both the performance and computational complexity of 6DMA channel estimation. Therefore, an alternative channel estimation approach is designed in the following for 6DMA THz systems, which exploits directional sparsity and the resulting spatial non-stationarities to achieve an optimal trade-off between performance and computational complexity, even when a large number of candidate antenna positions/rotations are involved.
\subsection{6DMA Surface-Wise Estimation}
Since the antennas on the same 6DMA surface experience similar channel gains, we treat each surface individually and localize the visible users on each surface employing a maximum likelihood (ML) algorithm.
Specifically, since the received noise, \(\mathbf{z}_m\), in \eqref{mo} is colored, it is necessary to perform a pre-whitening step. The noise covariance matrix is given by \(\mathbf{C}_m = \mathbb{E}[\mathbf{z}_m \mathbf{z}_m^H] = \sigma^2 \, \text{diag}(\mathbf{w}_{m,1}^H \mathbf{w}_{m,1}, \ldots, \mathbf{w}_{m,T}^H \mathbf{w}_{m,T})\). This matrix can be factorized using Cholesky decomposition as \(\mathbf{C}_m = \sigma^2 \mathbf{D}_m \mathbf{D}_m^H\), where \(\mathbf{D}_m \in \mathbb{C}^{T \times T}\) is a lower triangular matrix. Using this result, the whitened received signal becomes 
\begin{align}\label{6}
	\bar{\mathbf{y}}_m = \mathbf{D}_m^{-1} \mathbf{y}_m = \boldsymbol{\Gamma}_m \mathbf{h}_m(\mathbf{q}_m,\mathbf{u}_m) + \bar{\mathbf{z}}_m,
\end{align}
where $\boldsymbol{\Gamma}_m = \mathbf{D}_m^{-1} \mathbf{W}_m$ and $\bar{\mathbf{z}}_m = \mathbf{D}_m^{-1} \mathbf{z}_m$ is white noise within the signal bandwidth.

Based on \eqref{6}, we treat the 6DMA THz channel estimation problem as a parameter estimation problem. At the $m$-th 6DMA position-rotation pair, the ML estimate of \((d, \phi, \theta)\), denoted by $(\hat{d}^{(m)}, \hat{\phi}^{(m)},\hat{\theta}^{(m)})$, is given by
\begin{align}\label{mo3}
	&\!\!\!(\hat{d}^{(m)}, \hat{\phi}^{(m)},\hat{\theta}^{(m)})=\arg\min_{d,\phi,\theta} \nonumber\\
	& \| \bar{\mathbf{y}}_m\! -\! \nu\boldsymbol{\Gamma}_m{ \sqrt{g(\mathbf{q}_m,\mathbf{u}_m)}e^{-j\frac{2\pi}{\lambda}
			d_{m}}}\mathbf{a}(\mathbf{q}_m,
	\mathbf{u}_m) \|_2^2.
\end{align}
Following the grid search approach in \cite{loww}, the coarse search grid \(\Xi_{\mathrm{H}}\) is constructed as:
\begin{align}
	\Xi_{\mathrm{H}} = \{(d,\phi,\theta)| d &= D_{\min}, D_{\min} + \Delta d_{\mathrm{H}}, \ldots, D_{\max}; \nonumber\\
	\phi& = -\pi, -\pi + \Delta \phi_{\mathrm{H}}, \ldots, \pi;\nonumber\\
	\theta& = -\frac{\pi}{2}, -\frac{\pi}{2} + \Delta \theta_{\mathrm{H}}, \ldots, \frac{\pi}{2} \},
\end{align}
where \(\Delta d_{\mathrm{H}}\), \(\Delta \phi_{\mathrm{H}}\), and \(\Delta \theta_{\mathrm{H}}\) are the grid steps for distance and angles, and \([D_{\min}, D_{\max}]\) defines the range of the user-BS distance. From \eqref{mo3}, we note that the objective function is a quadratic polynomial in \(\nu\). We therefore derive its closed-form solution. Specifically, taking the derivative of the objective function in \eqref{mo3} w.r.t. \(\nu\) and setting it to zero, we obtain its estimate as
\begin{align} \label{mo5}
	\hat{\nu} = \frac{
		(\boldsymbol{\Gamma}_m{ \sqrt{g(\mathbf{q}_m,\mathbf{u}_m)}e^{-j\frac{2\pi}{\lambda}
				d_{m}}} \mathbf{a}(\mathbf{q}_m, \mathbf{u}_m))^H \bar{\mathbf{y}}_m}{
		\| \boldsymbol{\Gamma}_m{ \sqrt{g(\mathbf{q}_m,\mathbf{u}_m)}e^{-j\frac{2\pi}{\lambda}
				d_{m}}} \mathbf{a}(\mathbf{q}_m, \mathbf{u}_m) \|_2^2}.
\end{align}
By substituting \eqref{mo5} into \eqref{mo3}, the optimization problem in \eqref{mo3} can be reformulated as
\begin{align}\label{hu8}
&(\hat{d}^{(m)}, \hat{\phi}^{(m)}, \hat{\theta}^{(m)}) = \arg\max_{(d,\phi,\theta) \in \Xi_{\mathrm{H}}} \nonumber\\
& \frac{| (\boldsymbol{\Gamma}_m \sqrt{g(\mathbf{q}_m,\mathbf{u}_m)} e^{-j\frac{2\pi}{\lambda}d_{m}} \mathbf{a}(\mathbf{q}_m, \mathbf{u}_m))^H \bar{\mathbf{y}}_m |^2}{\| \boldsymbol{\Gamma}_m \sqrt{g(\mathbf{q}_m,\mathbf{u}_m)} e^{-j\frac{2\pi}{\lambda}d_{m}} \mathbf{a}(\mathbf{q}_m, \mathbf{u}_m) \|_2^2}.
\end{align}

\subsection{Multi-Surface Estimation Refinement}
After performing surface-wise estimation, the surfaces then collaborate to further refine channel estimation, thereby enhancing estimation accuracy. However, due to directional sparsity, the channel estimation derived from certain 6DMA position-rotation pairs may be inaccurate. To avoid degraded performance, we discard unreliable estimates from 6DMA position-rotation pairs that do not receive sufficient (or any) power from the user. To this end, a distance-based clustering method in \cite{diss} is adopted to obtain the final refined estimates of \((d, \phi, \theta, \nu)\), denoted by \((\bar{d}, \bar{\phi}, \bar{\theta}, \bar{\nu})\), from \(\{(\hat{d}^{(m)}, \hat{\phi}^{(m)}, \hat{\theta}^{(m)})\}_{m=1}^M\). Specifically, once \(\{\hat{d}^{(m)}, \hat{\phi}^{(m)}, \hat{\theta}^{(m)}\}_{m=1}^M\) is obtained, each triple \((\hat{d}^{(m)}, \hat{\phi}^{(m)}, \hat{\theta}^{(m)})\) is converted into its corresponding Cartesian coordinate, yielding 
\begin{align}
\{\hat{\mathbf{d}}_m=[\hat{x}_m, \hat{y}_m, \hat{z}_m]^T\}_{m=1}^M,
\end{align}
where \(\hat{\mathbf{d}}_m\) denotes the estimated user position derived from the \(m\)‑th 6DMA position‑rotation pair.
Then, a single cluster \(\mathcal{S}_1\) is first initialized by exploiting the estimate \(\hat{\mathbf{d}}_1\), with its center defined as \(\mathbf{c}_1 = \hat{\mathbf{d}}_1\). For each subsequent estimate \(\hat{\mathbf{d}}_m\), the Euclidean distance to all existing cluster centroids \(\mathbf{c}_i\), \(i = 1, \dots, N_c\), is computed as follows
\begin{align}
D^{(m,i)} = \|\hat{\mathbf{d}}_m - \mathbf{c}_i\|_2,
\end{align}
where \(N_c\) denotes the number of clusters formed up to the current iteration.
The index corresponding to the minimum distance is given by
\begin{align}\label{minn}
i^* = \arg\min_{1 \le i \le N_c} D^{(m,i)}.
\end{align}
If minimum distance \(D_{\min}^{(m,i^*)}\) satisfies \(D_{\min}^{(m,i^*)} \le \epsilon\) for predefined threshold \(\epsilon>0\), the estimate, \(\hat{\mathbf{d}}_m\), is added to cluster \(\mathcal{S}_{i^*}\) and the corresponding cluster center is updated as
\begin{align}\label{vv0}
\mathbf{c}_{i^*} \leftarrow \frac{1}{\lvert \mathcal{S}_{i^*}\rvert_\mathrm{c}}\sum_{i \in \mathcal{S}_{i^*}} \hat{\mathbf{d}}_i.
\end{align}
Conversely, if \(D_{\min}^{(m,i^*)} > \epsilon\), \(\hat{\mathbf{d}}_m\) is considered too distant from all existing clusters, prompting the creation of new cluster \(\mathcal{S}_{N_c+1}=\{m\}\) with \(\mathbf{c}_{N_c+1} =\hat{\mathbf{d}}_m\). After processing estimates from all 6DMA position-rotation pairs, the cluster index with the largest cardinality is identified as
\begin{align}\label{mmax}
m_{\max} = \arg\max_{1 \le i \le N_c} \lvert \mathcal{S}_i\rvert_\mathrm{c}.
\end{align}
Consequently, the cluster that contains the largest number of 6DMA position-rotation pairs is denoted by \(\mathcal{S}_{m_{\max}}\), where \(\lvert \mathcal{S}_{m_{\max}}\rvert_\mathrm{c} = S\). 

Since reliable estimates of the true user channel tend to concentrate within a single dense cluster, whereas inaccurate or noisy estimates are scattered among smaller clusters, 
the center of cluster $\mathcal{S}_{m_{\max}}$, denoted by \(\mathbf{c}_{m_{\max}}\), is designated as final user coordinate estimate \((\hat{x},\hat{y},\hat{z})\) (i.e., $(\hat{d}, \hat{\phi}, \hat{\theta})$ in polar coordinates). 
\begin{algorithm}[t]
	\caption{Directional-Sparsity-Driven 6DMA THz Channel Estimation Algorithm}
	\label{alg:6dma_estimation}
	\begin{algorithmic}[1]
		\STATE \textbf{Input:} $\Xi_{\mathrm{H}}$, $\Xi_{\mathrm{L}}$, $\epsilon$, $M$
		\STATE \textbf{Initialization:}  $\mathcal{S}_{1}=1$, $N_c=1$
		\STATE \emph{Step 1: 6DMA Surface-Wise Estimation}
        \STATE All 6DMA surfaces traverse \( M \) position-rotation pairs to collect channel measurements.
		\FOR{$m = 1$ to $M$}
		\STATE Obtain \((\hat{d}^{(m)}, \hat{\phi}^{(m)}, \hat{\theta}^{(m)})\) from \eqref{hu8} and \(\hat{\nu}\) from \eqref{mo5}.
		\ENDFOR
		\STATE \emph{Step 2: Multi-Surface Estimation Refinement}
		\STATE $\mathbf{c}_1=\hat{\mathbf{d}}_1$.
		\FOR{$m = 2$ to $M$}
		\STATE Determine the index $i^*$ according to \eqref{minn}.
		\IF{$D_{\min}^{(m,i^\star)} \leq \varepsilon$}
		\STATE $\mathcal{S}_{i^\star} = \mathcal{S}_{i^\star} \cup m$, and update $\mathbf{c}_{i^*}$ according to \eqref{vv0}.
		\ELSE
		\STATE $N_c=N_c+1$, $\mathcal{S}_{N_c} = \{m\}$ and \(\mathbf{c}_{N_c} =\hat{\mathbf{d}}_m\).
		\ENDIF
		\ENDFOR
		\STATE Identify $\mathcal{S}_{m_{\max}}$ according to \eqref{mmax}.
		\STATE Calculate $\bar{d}$, $\bar{\phi}$, $\bar{\theta}$, $\hat{\nu}$ via \eqref{mo1} and \eqref{mo12}.
		\STATE Reconstruct \(\hat{\mathbf{h}}$ at any position-rotation pair \((\mathbf{q}, \mathbf{u})\) via \eqref{go9}.
		\STATE \textbf{Return} $\bar{d}$, $\bar{\phi}$, $\bar{\theta}$, $\hat{\nu}$, and $\hat{\mathbf{h}}(\mathbf{q}, \mathbf{u})$.
	\end{algorithmic}
\end{algorithm}

Once coarse estimate $(\hat{d}, \hat{\phi}, \hat{\theta})$ is obtained, more accurate values, $(\bar{d}, \bar{\phi}, \bar{\theta})$, can be derived via multiple-surface-based joint channel estimation with smaller step sizes, i.e., $\Delta d_{\mathrm{L}}<\Delta d_{\mathrm{H}}$, $\Delta \phi_{\mathrm{L}}<\Delta \phi_{\mathrm{H}}$, and $\Delta \theta_{\mathrm{L}}<\Delta \theta_{\mathrm{H}}$ \cite{loww}. The corresponding fine search grid \(\Xi_{\mathrm{L}}\) is given by 
\begin{align}
	&\!\!\!\!\!\Xi_{\mathrm{L}} = \{(d,\phi,\theta)| \nonumber\\
	&\!\!\!\!\!d= \hat{d} - \overline{D}/2, \ldots, \hat{d} - \Delta d_{\mathrm{L}}, \hat{d}, \hat{d} + \Delta d_{\mathrm{L}}, \ldots, \hat{d} + \overline{D}/2; \nonumber\\
	&\!\!\!\!\!\phi = \hat{\phi} - \overline{\Phi}/2, \ldots, \hat{\phi} - \Delta \phi_{\mathrm{L}}, \hat{\phi}, \hat{\phi} + \Delta \phi_{\mathrm{L}}, \ldots, \hat{\phi} + \overline{\Phi}/2 ;\nonumber\\
		&\!\!\!\!\!\theta = \hat{\theta} - \overline{\Theta}/2, \ldots, \hat{\theta} - \Delta \theta_{\mathrm{L}}, \hat{\theta}, \hat{\theta} + \Delta \theta_{\mathrm{L}}, \ldots, \hat{\theta} + \overline{\Theta}/2 \},
\end{align}
where \(\overline{D}\), \(\overline{\Phi}\), and \(\overline{\Theta}\) define the refined distance and angular ranges of \(\Xi_{\mathrm{L}}\). Based on \(\Xi_{\mathrm{L}}\), the final parameter estimates are jointly determined by multiple 6DMA surfaces as follows:
\begin{subequations}
	\label{moww}
\begin{align}\label{mo1}
	(\bar{d}, \bar{\phi}, \bar{\theta}) &= \arg\max_{(d,\phi,\theta)\in \Xi_{\mathrm{L}}} \frac{\left| (\boldsymbol{\Gamma}_{\mathrm{L}} \mathbf{a}_{\mathrm{L}}(\mathbf{q},\mathbf{u}))^H \bar{\mathbf{y}}_{\mathrm{L}} \right|^2}{\left\| \boldsymbol{\Gamma}_{\mathrm{L}} \mathbf{a}_{\mathrm{L}}(\mathbf{q},\mathbf{u}) \right\|_2^2},\\
	\bar{\nu}&= \frac{ (\boldsymbol{\Gamma}_{\mathrm{L}} \mathbf{a}_{\mathrm{L}}(\mathbf{q},\mathbf{u}))^H \bar{\mathbf{y}}_{\mathrm{L}} }{\left\| \boldsymbol{\Gamma}_{\mathrm{L}} \mathbf{a}_{\mathrm{L}}(\mathbf{q},\mathbf{u}) \right\|_2^2},\label{mo12}
\end{align}
\end{subequations}
where $\boldsymbol{\Gamma}_{\mathrm{L}} = \text{blkdiag}(\boldsymbol{\Gamma}_{j_1}, \ldots, \boldsymbol{\Gamma}_{j_S})$, $\bar{\mathbf{y}}_{\mathrm{L}}\! =\! [\bar{\mathbf{y}}_{j_1}^T, \ldots, \bar{\mathbf{y}}_{j_S}^T]^T$, and $\mathbf{a}_{\mathrm{L}} = [\mathbf{a}_{j_1}^T{ \sqrt{g(\mathbf{q}_{j_1},\mathbf{u}_{j_1})}e^{-j\frac{2\pi}{\lambda}
		d_{j_1}}},  \ldots, 
	\mathbf{a}_{j_S}^T \sqrt{g(\mathbf{q}_{j_S},\mathbf{u}_{j_S})} \\e^{-j\frac{2\pi}{\lambda}
		d_{j_S}} ]^T$ denote the aggregated analog combining matrices, received signals, and steering vectors of 6DMA position-rotation pairs in set $\mathcal{S}_{\max}$, respectively. Here, \(j_1, j_2, \dots, j_S\) are the indices of the 6DMA position-rotation pairs that belong to the cluster \(\mathcal{S}_{m_{\max}}\). 

Given $\bar{d}$, $\bar{\phi}$, $\bar{\theta}$, and $\bar{\nu}$, the channel between the user and a 6DMA surface at an arbitrary position-rotation pair \((\mathbf{q}, \mathbf{u})\) can be reconstructed according to \eqref{ap} as
\begin{align}\label{go9}
\hat{\mathbf{h}}(\mathbf{q}, \mathbf{u}) = \mathbf h(\mathbf q,\mathbf u)
\bigl\lvert_{d=\bar d,\;\phi=\bar\phi,\;\theta=\bar\theta,\;\nu=\bar\nu}.
\end{align}

The main steps of the proposed channel estimation procedure are summarized in Algorithm 1. The complexity of the surface-wise estimation and the multi-surface refinement estimation is \(\mathcal{O}(TNB|\Xi_{\mathrm{H}}|_{\mathrm{c}}+B{N_c})\) and \(\mathcal{O}(TSN|\Xi_{\mathrm{L}}|_{\mathrm{c}})\), respectively. The proposed channel estimation approach reduces complexity by utilizing surface-based structures and incorporating only the reliable estimates from 6DMA position-rotation pairs.
\section{Simulation Results}
	\vspace{-1 mm}
In the simulation, we set \(B = 8\), \(N = 16\), \(K = 25\), \(T = 10\), the antenna spacing to \(\frac{\lambda}{2}\), the carrier frequency to 0.1 THz, and \(\mathcal{C}\) as a cube with side length \(A = 0.5\) meters, unless otherwise specified. The array aperture at 6DMA-BS can be determined by $D=A\sqrt{3}$ m.  Based on the definition of Rayleigh distance (RD), given by \(\mathrm{RD} = \frac{2D^2}{\lambda}\) \cite{liuy}, the boundary separating the near-field and far-field regions in our simulation setup is 500 m.
Users are randomly located within a 3D coverage area defined by a spherical annulus with radial distances ranging from 20 m to 800 m from BS center. We set $\Delta d_{\mathrm{H}} = \lambda$, $\Delta \theta_{\mathrm{H}} =\Delta \phi_{\mathrm{H}}= \frac{\pi}{1000}$, $\Delta d_{\mathrm{L}} = 0.025\lambda$, $\Delta \theta_{\mathrm{L}}=\Delta \theta_{\mathrm{H}} = \frac{\pi}{5000}$. The directive effective antenna gain for the 6DMA-enabled BS follows the 3GPP standard \cite{shao20246d}. 

We assume that \( M \) candidate position-rotation pairs are evenly distributed across the largest spherical surface that fits within the 6DMA-BS site space \(\mathcal{C}\). For performance comparison, we consider the least squares (LS)-based estimation scheme as the baseline. In this scheme, an LS channel estimate is first derived from the received data of $M$ 6DMA position-rotation pairs, and then a grid search over distance and angles is performed based on the channel estimate to extract \((d, \phi, \theta, \nu)\). 
\begin{figure*}[htbp]
	\begin{minipage}[t]{0.27\linewidth}
		\centering
		\includegraphics[width=2.56in]{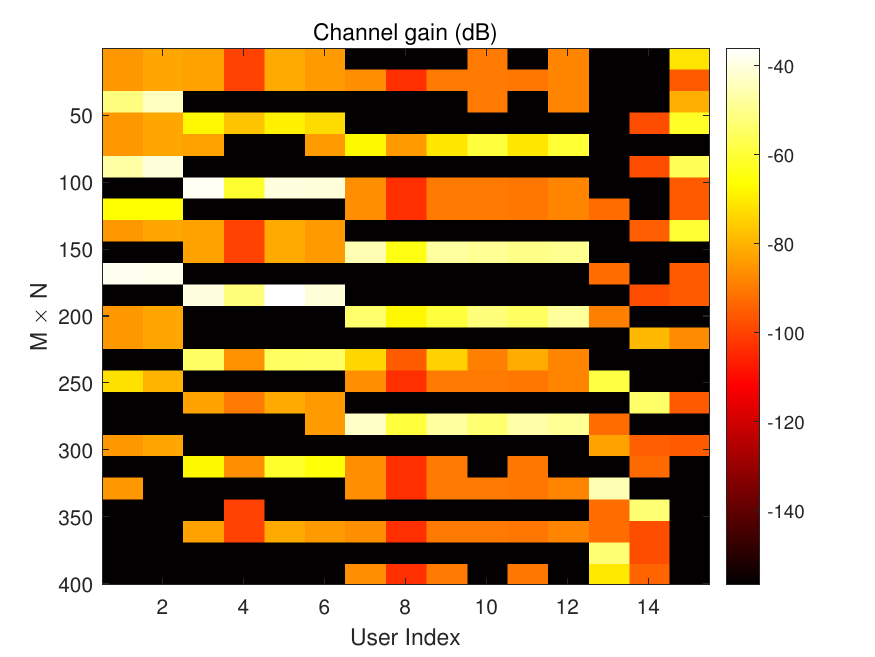}
		\setlength{\abovecaptionskip}{-15pt}
		\setlength{\belowcaptionskip}{-15pt}
		\caption{An illustration of the directional sparsity in the LoS 6DMA channel.}
		\label{heatmap}
	\end{minipage}%
	\hspace{.45in}
	\begin{minipage}[t]{0.30\linewidth}
		\centering
		\includegraphics[width=2.560in]{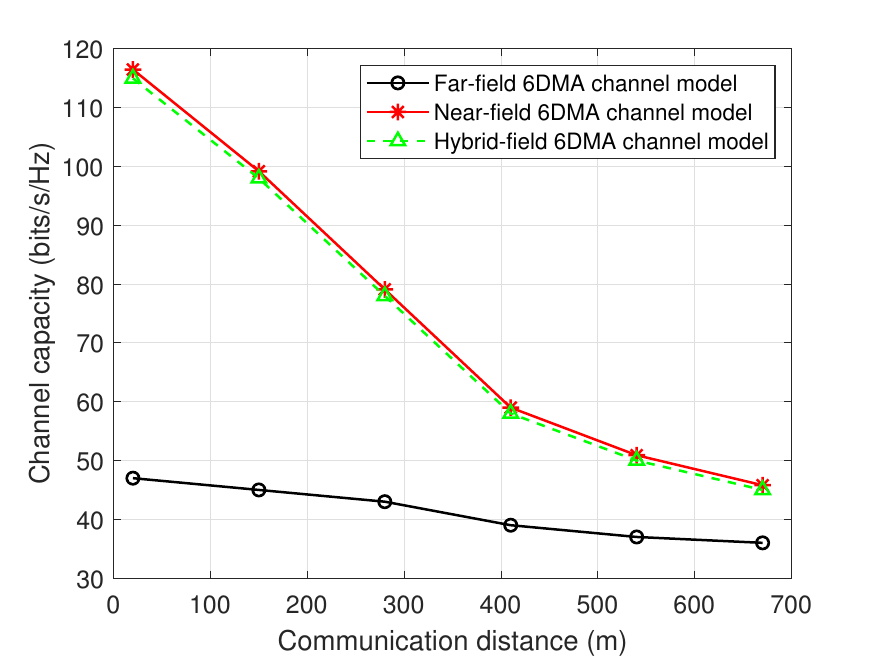}
		\setlength{\abovecaptionskip}{-8pt}
		\setlength{\belowcaptionskip}{-15pt}
		\caption{Channel capacity with various communication distances.}
		\label{nearfar}
	\end{minipage}%
	\hspace{.25in}
	\begin{minipage}[t]{0.27\linewidth}
		\centering
		\includegraphics[width=2.560in]{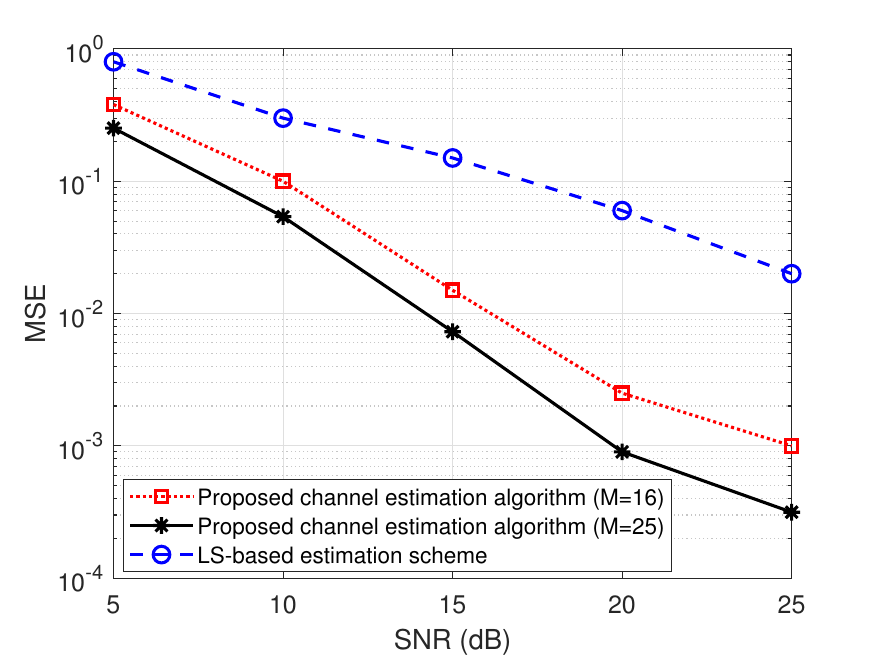}
		\setlength{\abovecaptionskip}{-15pt}
		\setlength{\belowcaptionskip}{-15pt}
		\caption{The MSE of 6DMA THz channel estimation versus SNR.}
		\label{snrc}
	\end{minipage}
	\vspace{-0.69cm}
\end{figure*}
First, Fig. \ref{heatmap} illustrates the directional sparsity of the 6DMA channels from the \(K\) users to the \(M\) measured 6DMA position rotation pairs. It is observed that users are served by only a limited number of 6DMA position rotation pairs. In addition, the block sparsity indicates that a 6DMA surface is the smallest unit that exhibits stationarity, that is, an even channel power distribution across antennas on the same 6DMA surface.

The channel capacity over varying communication distances between the BS and the user is shown in Fig.~\ref{nearfar}. To compute capacity, the positions and rotations of \(B\) 6DMA surfaces are randomly selected from \(M\) available 6DMA position-rotation pairs. The results indicate that the average sum rate achieved by the hybrid-field 6DMA THz channel model is close to that of the near-field model and remains considerably higher than the far-field counterpart. This is because the far-field model ignores spherical-wave effects and fails to accurately capture channel characteristics in the near-field transmission. In contrast, the hybrid-field channel model retains near-field accuracy with fewer parameters. When the communication distance exceeds the Rayleigh distance of 500 meters, the capacities derived from all three models become similar. Hence, the hybrid-field 6DMA THz channel model provides a reliable and efficient solution across both near-field and far-field transmission regimes.

To evaluate the channel reconstruction performance, Fig. \ref{snrc} shows the mean MSE of channel estimation versus the signal-to-noise ratio (SNR) for the proposed channel estimation algorithm and LS scheme. We observe that the proposed algorithm yields a significantly lower MSE compared to the LS-based method. This is because the proposed approach uses clustering to remove outliers and highly erroneous estimates caused by directional sparsity and retains only the most representative cluster for more accurate user localization. The LS‑based scheme does not exploit this clustering step. Moreover, increasing the number of candidate position-rotation pairs further improves channel estimation accuracy, since more 6DMA surfaces capture the user’s signals from different directions.
		
\section{Conclusion}
In this work, we have proposed a novel hybrid-field channel model to accurately represent the 6DMA channel in THz systems with limited channel characteristics. We then proposed a low-complexity channel estimation algorithm by taking advantage of the non-stationary premise that not all users are served by all 6DMA position-rotation pairs. Simulation results showed that the proposed hybrid-field channel model yielded a capacity close to the ground-truth near-field model with low complexity, and the designed channel estimation algorithm achieved significantly lower MSE than existing approaches. For future work, we will focus on optimizing the antenna positions and rotations in the 6DMA THz system through hybrid-field 6DMA channel modeling.

	\bibliographystyle{IEEEtran}
	\bibliography{fabs}
\end{document}